%% file: SamplePaths.tex
\documentclass[11pt]{article}
\pdfoutput=1
\usepackage[margin=2.7cm]{geometry} 
\usepackage[english]{babel}
\usepackage{graphicx,subfigure}
\usepackage{float}
\usepackage{amsmath}
\usepackage{subeqnarray}
\begin{document}
\newcommand{\ee}{\`e\ }
\renewcommand{\aa}{\`a\ }
\newcommand{\oo}{\`o\ }
\newcommand{\uu}{\`u\ }
\newcommand{\ket}[1]{\ensuremath {|\: #1 \: \rangle}}
\newcommand{\bra}[1]{\ensuremath{\langle \: #1 \:|}}
\newcommand{\braket}[2]{\ensuremath{\langle \: #1 \: | \: #2 \: \rangle}}
\newcommand{\ketbra}[2]{\ensuremath{| \: #1 \:\rangle \langle \: #2 \:  |}}
\newcommand{\ves}[2]{\ensuremath{#1_1,#1_2, \ldots, #1_{#2}}}
\newcommand{\ve}[2]{\ensuremath{#1(1),#1(2) \ldots, #1(#2)}}
\newcommand{\Hc}{\ensuremath{\mathcal{H}_{cursor\;}}}
\newcommand{\Hr}{\ensuremath{\mathcal{H}_{register\;}}}
\newcommand{\Hcr}{\ensuremath{\mathcal{K}_{cur\;}}}
\newcommand{\Hrr}{\ensuremath{\mathcal{K}_{reg\;}}}
\newcommand{\Hm}{\ensuremath{\mathcal{H}_{machine\;}}}
\newcommand{\eref}[1]{(\ref{#1})}
\newcommand{\fref}[1]{figure \ref{#1}}
\newcommand{\sref}[1]{section \ref{#1}}

\markboth{de Falco, D. and Tamascelli D.}
{Dynamical kickback and non commuting impurities in a spin chain}

%
\title {Dynamical kickback and noncommuting impurities in a spin chain}
\author{Diego de Falco$^{1,2}$ and Dario Tamascelli$^{1,2}$\\ 
$^1$ Dipartimento di Scienze dell'Informazione, Universit\aa degli Studi di Milano, \\
via Comelico 39, 20135 Milano, Italy,\\[3pt]
$^2$ CIMAINA, Centro Interdipartimentale Materiali e Interfacce Nanostrutturati,\\ Universit\aa degli Studi di Milano\\[3pt]
e-mail: \texttt{defalco@dsi.unimi.it, tamascelli@dsi.unimi.it}}
%

%
\maketitle
%
%
\begin{abstract}
In an interacting continuous time quantum walk, while the walker (the cursor) is moving on a graph, computational primitives (unitary operators associated with the edges) are applied to ancillary qubits (the register). The model with one walker was originally proposed by R. Feynman, who thus anticipated many features of the Continuous Time Quantum Walk (CTWQ) computing paradigm. In this note we examine the behaviour of an  interacting CTQW with two walkers and examine the interaction of the walkers with noncommuting primitives. We endow such a walk with a notion of trajectory, in the sense of sample path of an associated Markov process,  in order to use such notions as sojourn time and first passage time as heuristic tools for gaining intuition about its behaviour.\\[3pt]
\textbf{Keywords:} Continuous time quantum walks; birth and death processes; quantum annealing.
\end{abstract}
\input{introduction.tex}
\input{Kickback.tex}
\input{sample.tex}
\input{conclusionsandoutlook.tex}
\newpage
\bibliography{mybib}
\bibliographystyle{unsrt}
\end{document}

%% file: introduction.tex
\section{Introduction}
We consider a collection of spin $1/2$ systems $\underline{\tau}(j)=(\tau_1(j),\tau_2(j),\tau_3(j))$, \mbox{$j \in \Lambda_s \equiv \{1,2,\ldots,s\}$}, coupled with an \emph{ancilla} qubit $\underline{\sigma}=\{\sigma_1,\sigma_2,\sigma_3\}$ through a Hamiltonian of the form:
\begin{equation}
 H(a,b)=- \frac{1}{2} \sum_{x=1}^{s-1} U_x \otimes \tau_+(x+1)\  \tau_-(x) + U_x^{-1} \otimes \tau_+(x) \ \tau_-(x+1),
\end{equation}
where $\tau_{\pm}(j)=(\tau_1(j) \pm i \tau_2(j))/2$. The integers $a,b$ are supposed to satisfy $1<a<b<s$; the unitary operators $U_x$ act on the state space of the ancilla qubit. We will take, in this note,
\begin{equation} \label{eq:operatori}
 U_a=\sigma_1,\ U_b=\sigma_3
\end{equation}
and will suppose that all the remaining $U_x$ are the identity operator. We will consider an initial condition in the eigenspace belonging to the eigenvalue 2 of the conserved \emph{number operator}
\begin{equation}
 N_3= \sum_{x=1}^s \frac{1+\tau_3(x)}{2},
\end{equation}
and we will refer the system to the orthonormal basis \ket{(x_1,x_2),\zeta}, where $\zeta \in \{-1,1\}$ and $1 \leq x_1 <x_2 \leq s$, formed by the simultaneous eigenstates of $\tau_3(x)$, $x\in \Lambda_s$, and $\sigma_3$, belonging, respectively, to the eigenvalue $+1$ of $\tau_3(x_1)$ and $\tau_3(x_2)$, to the eigenvalue $-1$ of the remaining $\tau_3(x)$ and to the eigenvalue $+1$ of $\zeta$.\\
We will look at the above system from two points of view:
\begin{enumerate}
\def\theenumi{\roman{enumi}}
\item \label{it:anderson} as an Anderson model \cite{lee85} with noise on the hopping parameters relative to the links $\{a,a+1\}$ and $\{b,b+1\}$, with the peculiarity that the ``random values'' of these parameters are determined by the non commuting observables $\sigma_1$ and $\sigma_3$;
\item \label{it:feynman}as a version of Feynman's model of a quantum computer \cite{feyn86}, where the motion of spin-up excitations of the $\tau$ field (the \emph{clock}) administers the primitives $U_a$ and $U_b$ to the ancilla (the \emph{register}).
\end{enumerate}
For an extensive analysis of related models in the subspace belonging to the eigenvalue $1$ of $N_3$ we refer the reader to  \cite{defa06b}. In the $N_3=1$ subspace (because of Peres' conservation laws \cite{peres85}) the presence of an ancilla cannot affect the motion of the clocking excitation. In the $N_3=2$ subspace we will, on the contrary, give evidence of a peculiar three-body effect involving the two clocking excitations and the ancilla qubit, related to the fact that $U_a$ and $U_b$ do not commute. 

%% file: Kickback.tex
\section{Dynamical kickback} \label{sec:kickback}
\begin{figure}[t]
	\centering
		\subfigure[]{\label{fig:figura1a} \includegraphics[width=5cm]{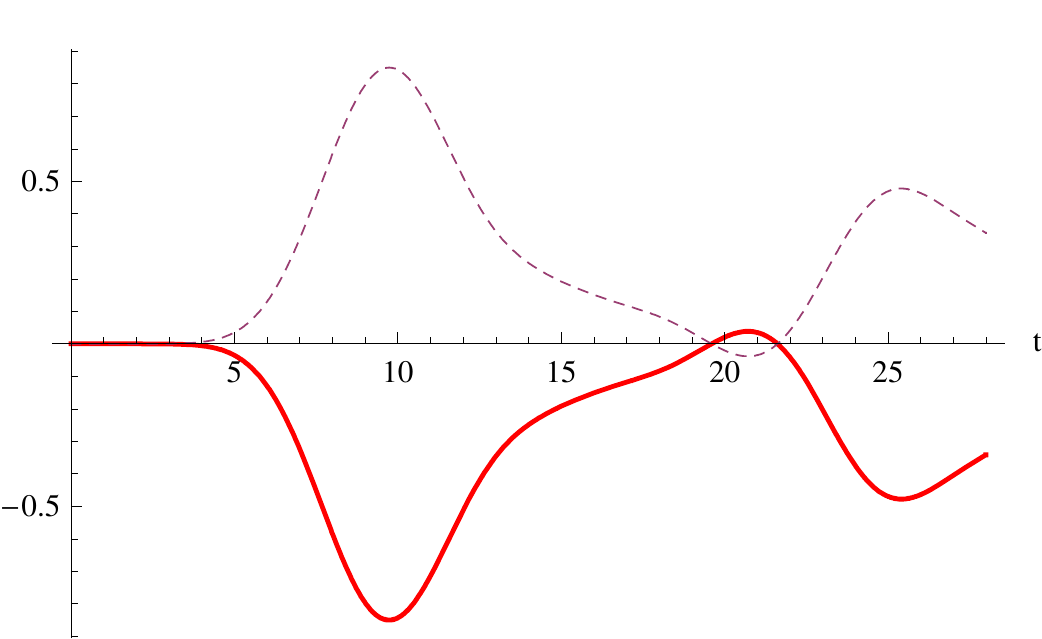}} 
		\hspace{2cm}
		\subfigure[]{\label{fig:figura1b} \includegraphics[width=5cm]{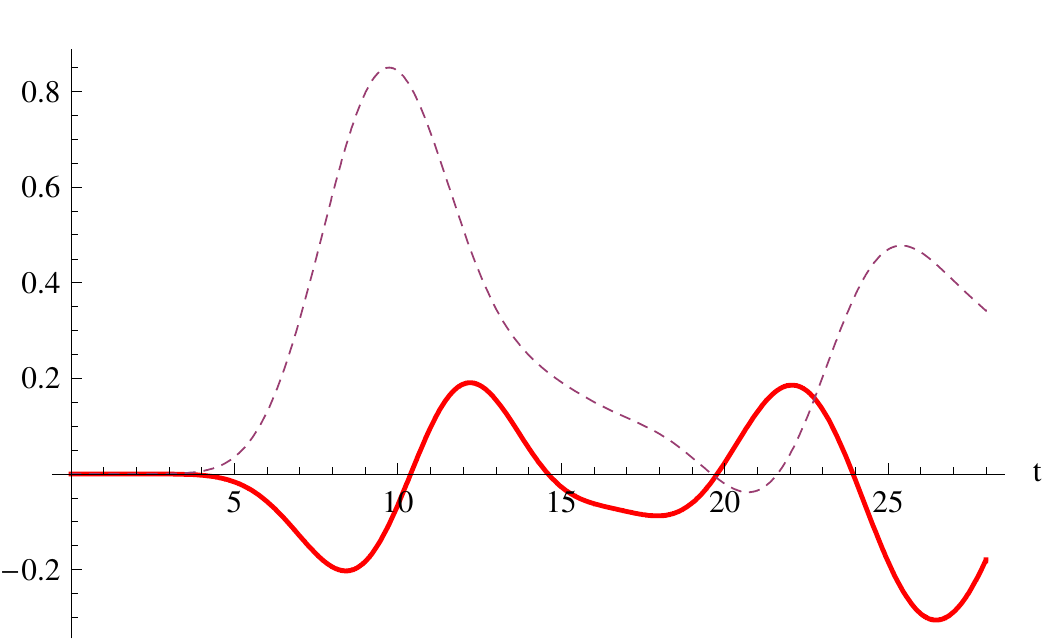}} 
	\caption{Frame (a): $s=7,\ a=4, \ b=5$. Frame (b): $s=7,\ a=3, \ b=5$. Solid lines: the probability amplitude $-\psi_t((s-1,s),1)$ as a function of time, under the initial condition \eref{eq:ciperla}. The dashed lines refer to the free case.}
	\label{fig:figura1}
\end{figure}

The study of the system introduced in the previous section is made easy by the fact that the two following projectors are constants of motion:
\begin{eqnarray}
 P_\pm  =   \sum_{1 \leq x_1 < x_2 \leq s} & &  \ket{(x_1,x_2),\pm(-1)^{\theta(x_1-a)+\theta(x_2-a)}} \nonumber \\
& & \bra{(x_1,x_2),\pm(-1)^{\theta(x_1-a)+\theta(x_2-a)}},
\end{eqnarray}
where $\theta(x)=$ \emph{if} $x>0$ \emph{then} 1, \emph{else} 0. If, as we will always do in this note, we consider the evolution of the system from the initial state
\begin{equation} \label{eq:ciperla}
 \ket{\psi_0}= \ket{(1,2),+1},
\end{equation}
we will be interested only in the matrix elements of the Hamiltonian $H(a,b)$ between states belonging to the range of $P_+$:
 \begin{align} \label{eq:hamiltridotta}
& h_+((x_1,x_2),(y_1,y_2)) \equiv  \nonumber \\
&\equiv \bra{(x_1,x_2),(-1)^{\theta(x_1-a)+\theta(x_2-a)}} H(a,b) \ket{(y_1,y_2),(-1)^{\theta(y_1-a)+\theta(y_2-a)}} = \nonumber \\
&= -\frac{1}{2} \left ( \delta_{x_1,y_1} (\delta_{x_2,y_2+1}+\delta_{x_2,y_2-1})  + 
\delta_{x_2,y_2} (\delta_{x_1,y_1+1} + \delta_{x_1,y_1-1}) \right ) + \nonumber \\
&+ (1-\theta(x_1-a))\delta_{x_1,y_1} (\delta_{x_2,b}\  \delta_{y_2,b+1} + \delta_{x_2,b+1} \ \delta_{y_2,b}).
\end{align}
Looking at the model from the point of view (\ref{it:feynman}), it performs quite a trivial computational task: starting with the two clocking excitations in positions 1 and 2 and with the register ``up'', it returns, \emph{if} the two clocking excitations are found in the terminal positions $s-1$ and $s$, the register again ``up''.\\
In the case $b=a+1$ the only track of the fact that, in applying the identity to the register, the machine has temporarily flipped it (by applying $\sigma_1$), can be seen by comparing the corresponding probability amplitude with the one for the free case in which all the $U_x$ are the identity operator: the inversion of phase (``dynamical kickback'') shown in figure \fref{fig:figura1a} is easily understood  by thinking that, while the two clocking excitations move to the right, they restore the register into the ``up'' state by applying \emph{minus} the identity operator to the register in the successive steps (from right to left) $\sigma_3 \sigma_1 \sigma_3 \sigma_1$.\\
%
%
\begin{figure}[t]
	\centering
		\subfigure[]{\label{fig:figura2a} \includegraphics[width=4.5cm]{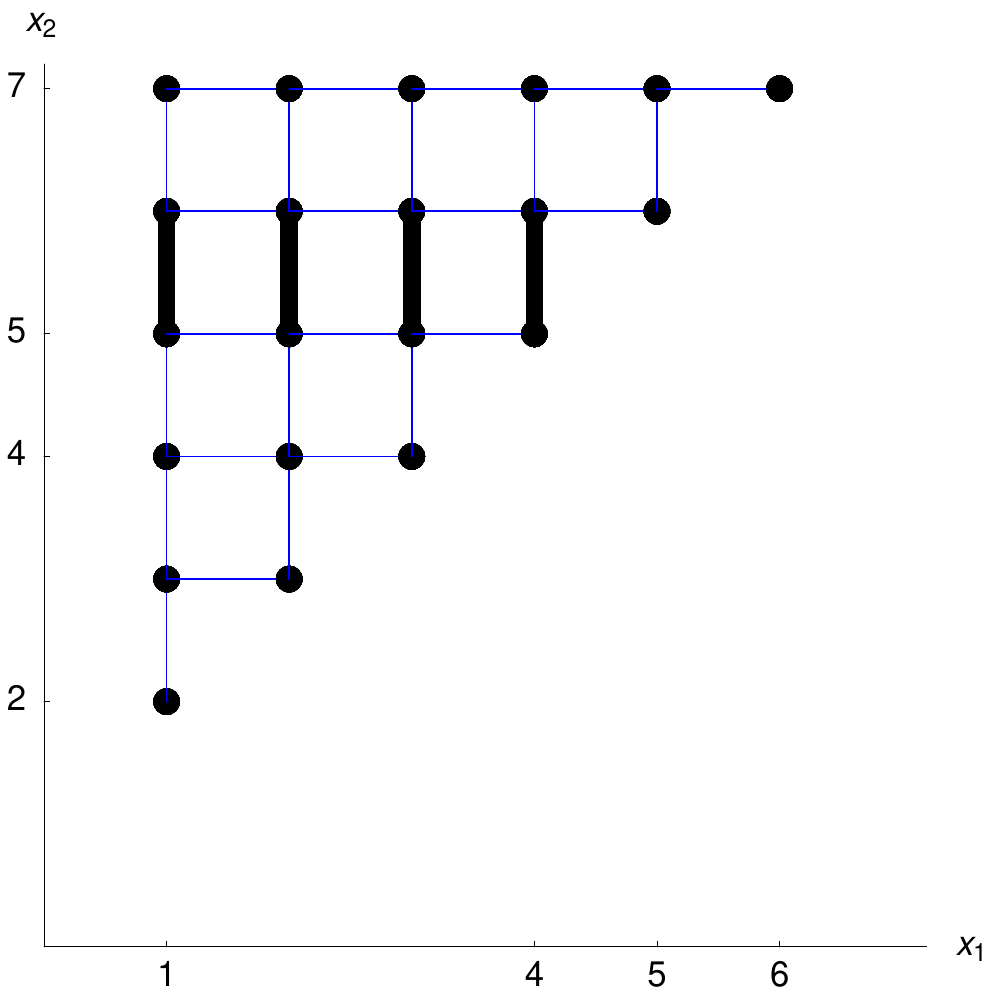}} 
		\hspace{2cm}
		\subfigure[]{\label{fig:figura2b} \includegraphics[width=4.5cm]{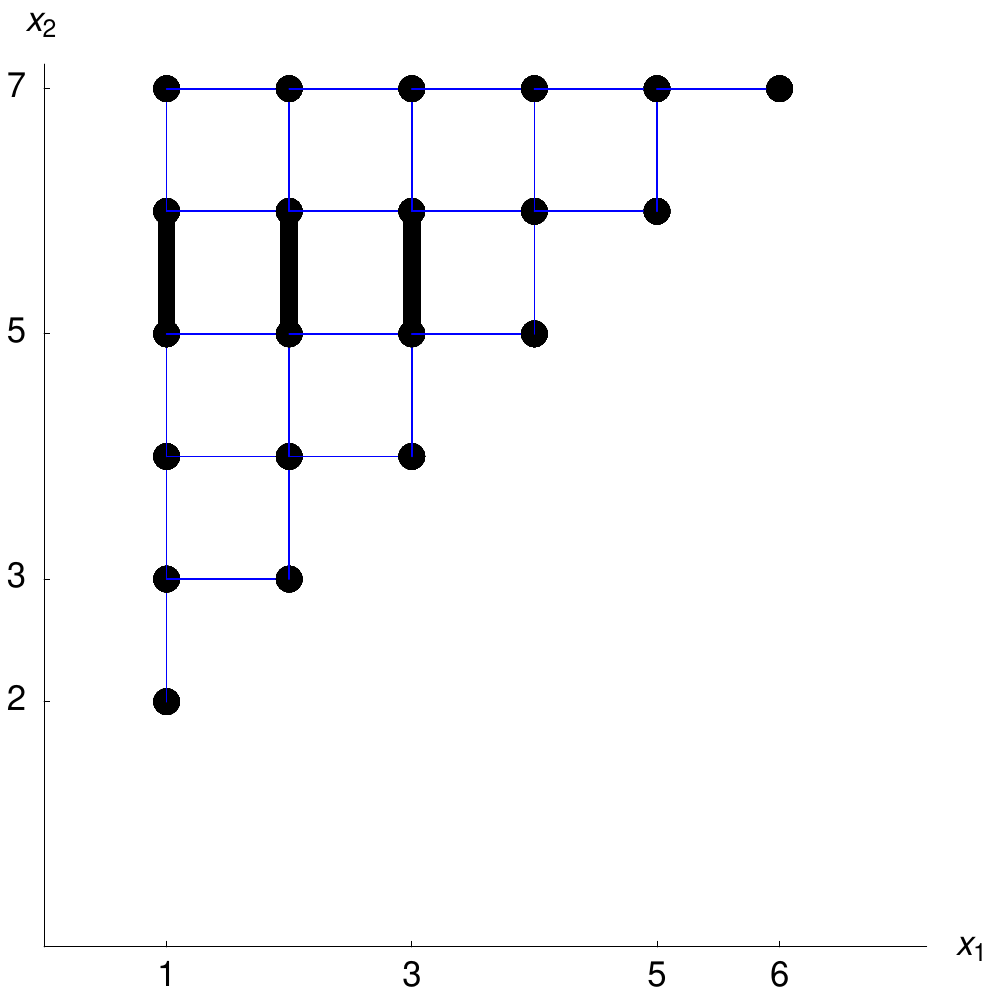}} 
	\caption{ Graphical representation of the Hamiltonian \eref{eq:hamiltridotta} on the weighted graph having the set of vertices $\{ (x_1,x_2) \in \Lambda_s \times \Lambda_s, 1 \leq x_1 < x_2 \leq s \}$ with edges between nearest neighbour sites. Edges to which \eref{eq:hamiltridotta} attributes a positive weight are represented by thick lines. \mbox{Frame (a): $s=7,\ a=4, \ b=a+1=5$. Frame (b): $s=7,\ a=3,$ \mbox{$b=a+2=5$}}.}
	\label{fig:figura2}
\end{figure}
\noindent The above simplistic description of the evolution of the ``computation'' with two clocking excitations holds only in the particular case $b=a+1$ considered up to now, as made clear by the example with $b=a+2$ shown in \fref{fig:figura1b}.\\
By direct inspection of the Hamiltonian \eref{eq:hamiltridotta} and, in particular, of the \emph{weighted} graph on which our quantum walk takes place (figure 2), the role of the positions $a$ and $b$ of the two noncommuting impurities $U_a=\sigma_1$ and $U_b=\sigma_3$ is easily understood in the context of an interference phenomenon. The term 
\[
(1-\theta(x_1-a)) \ \delta_{x_1, y_1} \ (\delta_{x_1,b} \ \delta_{y_2,b+1} + \delta_{x_2,b+1} \ \delta_{y_2,b}) 
\]
in $h_+((x_1,x_2), (y_1,y_2))$ shows that in the situation $b=a+1$ of \fref{fig:figura1a} \emph{all} the amplitudes $\psi_t(x_1,b+1)$, $1 \leq x_1 \leq b$ are phase inverted with respect to the free case evolving according to the finite difference Laplacian
\[
  \delta_{x_1, y_1} \ (\delta_{x_2,y_2+1} + \delta_{x_2,y_2-1}) + \delta_{x_2,y_2} (\delta_{x_1,y_1+1} + \delta_{x_1,y_1-1})).
\]
In the situation $b=a+2$ of \fref{fig:figura1b}, on the contrary, \emph{only} the amplitudes $\psi_t(x_1,b+1)$, $1 \leq x_1 \leq a$ are phase inverted and deviations with respect to the free case take place because of interference with the uninverted signal $\psi_t(a+1,b+1)$: that this interference can be destructive, suppressing the probability of the two excitations ever going beyond the noncommuting impurities, is shown by comparison of the two frames of \fref{fig:figura1}. A notational remark: for a wave function $\psi_t((x_1,x_2),\zeta)$ in the range of $P_+$ we are suppressing explicit indication of \mbox{the argument $\zeta$.}

%% file: sample.tex
\section{Sample paths} \label{sec:sample}
\begin{figure}[h]
 \centering
 \subfigure[]{\label{fig:figuraX1comp} \includegraphics[width=5.5cm]{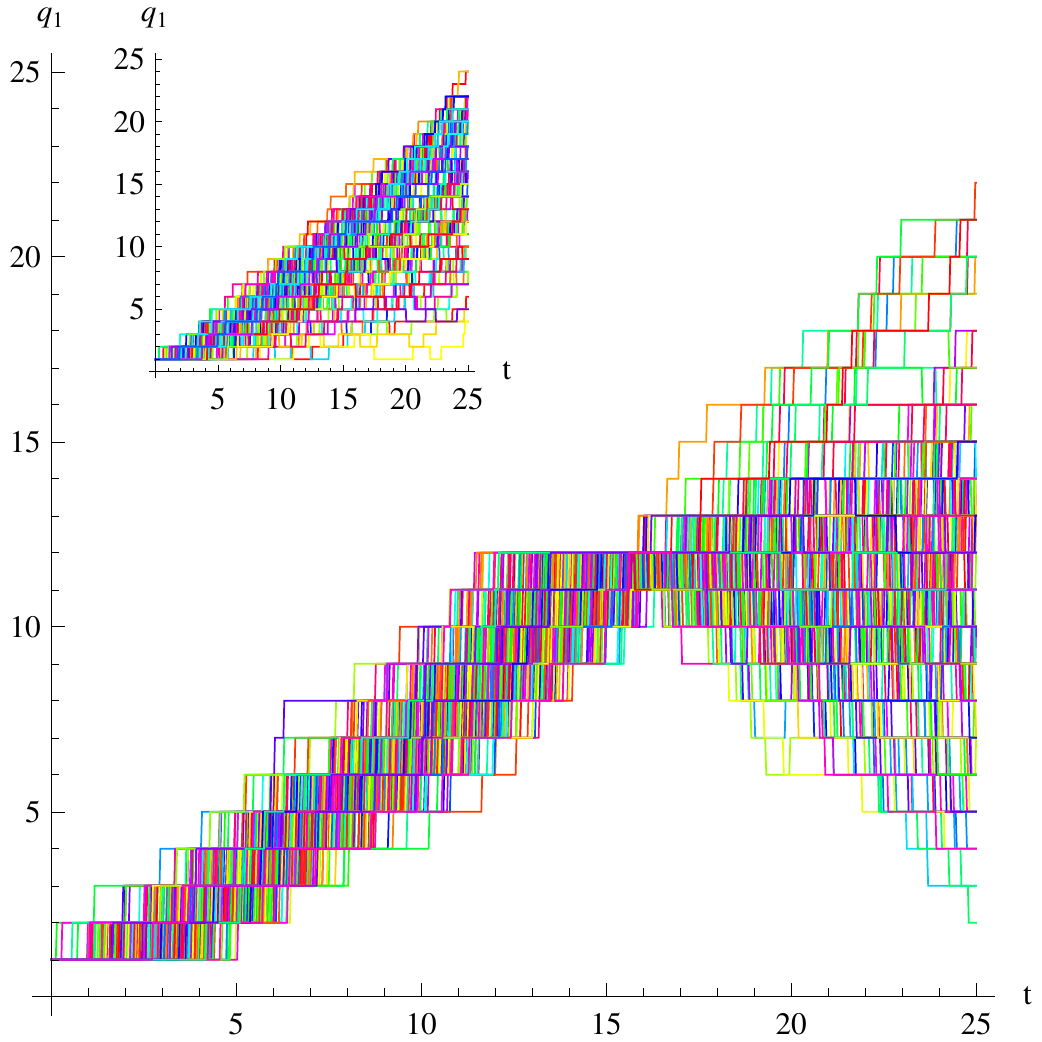}}
\hspace{1cm}
\subfigure[]{\label{fig:figuraX2comp} \includegraphics[width=5.5cm]{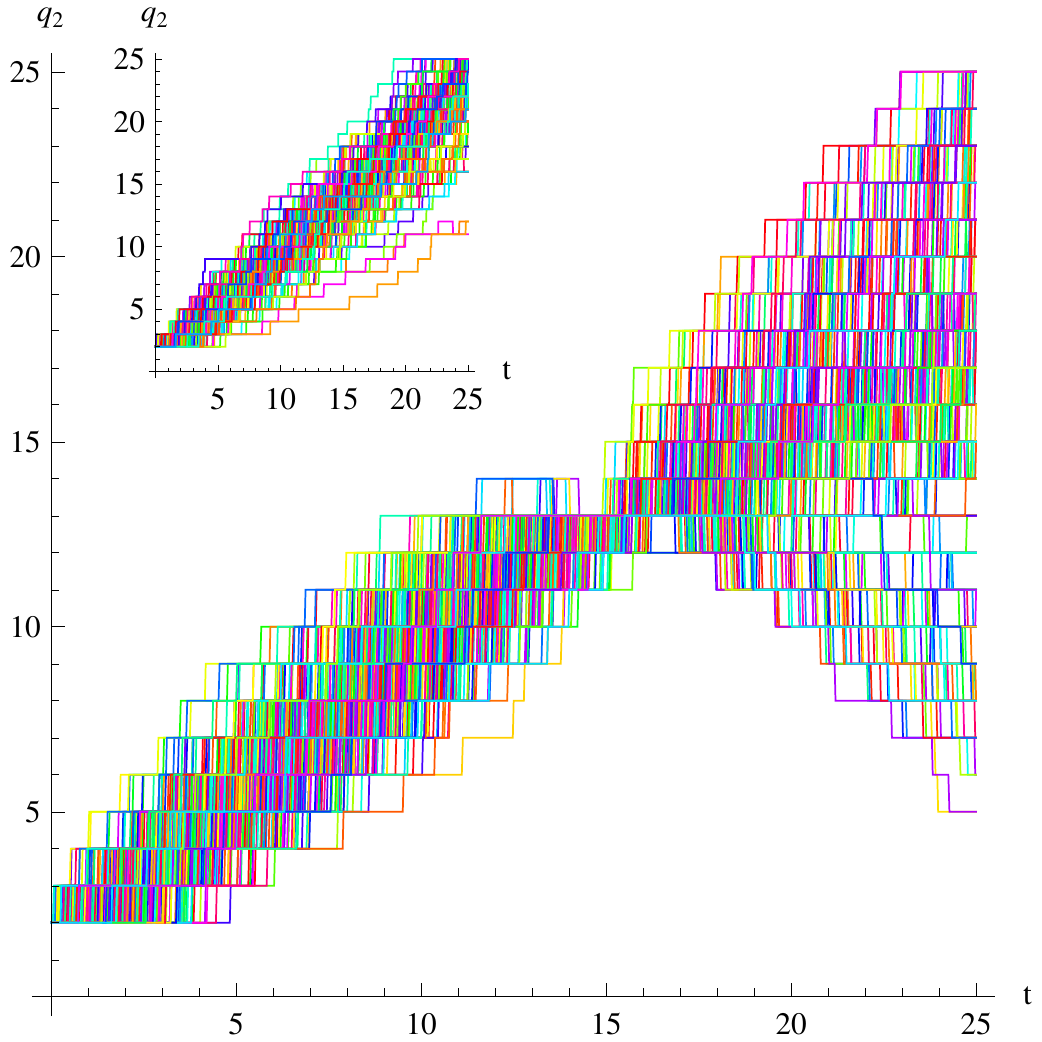}}
 \label{fig:Xcomp}
\caption{ $s = 25,\ a =11,\ b = a + 2 =13$. Frames (a) and (b):  sample paths of the interacting processes $q_1(t)$ and $q_2(t)$ respectively. Only the trajectories, in our sample of size $10^4$, that hit the site $(a +1,b)$ are shown. In the insets, the corresponding trajectories of the free process $(q_1^0(t),q_2^0(t))$ are shown for comparison purposes.}
\end{figure}

\begin{figure}[h]
 \centering
 \subfigure[]{\label{fig:figura27} \includegraphics[width=4.5cm]{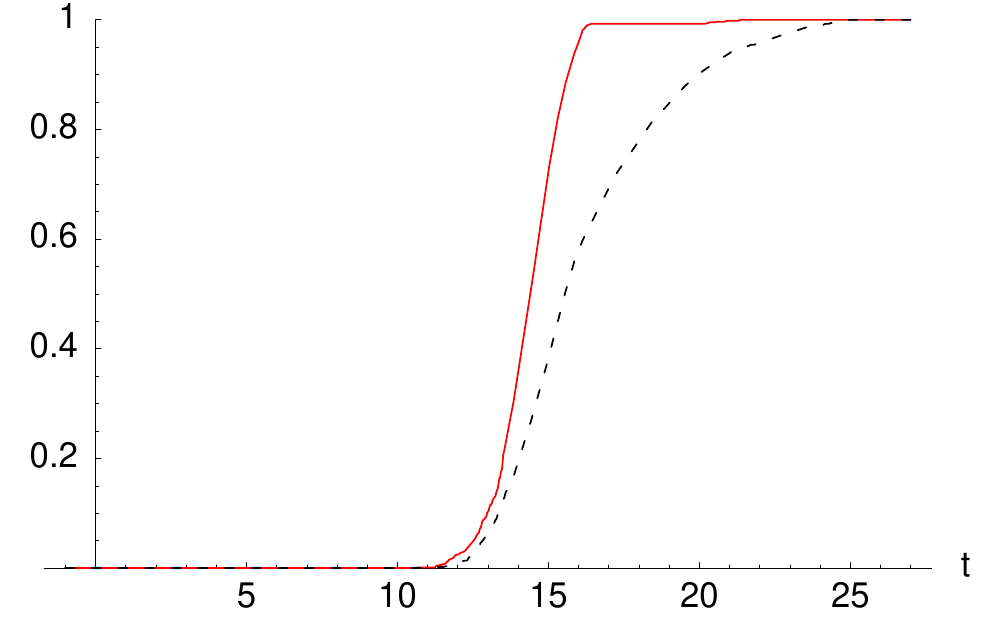}}
\hspace{2cm}
\subfigure[]{\label{fig:figura24} \includegraphics[width=4.5cm]{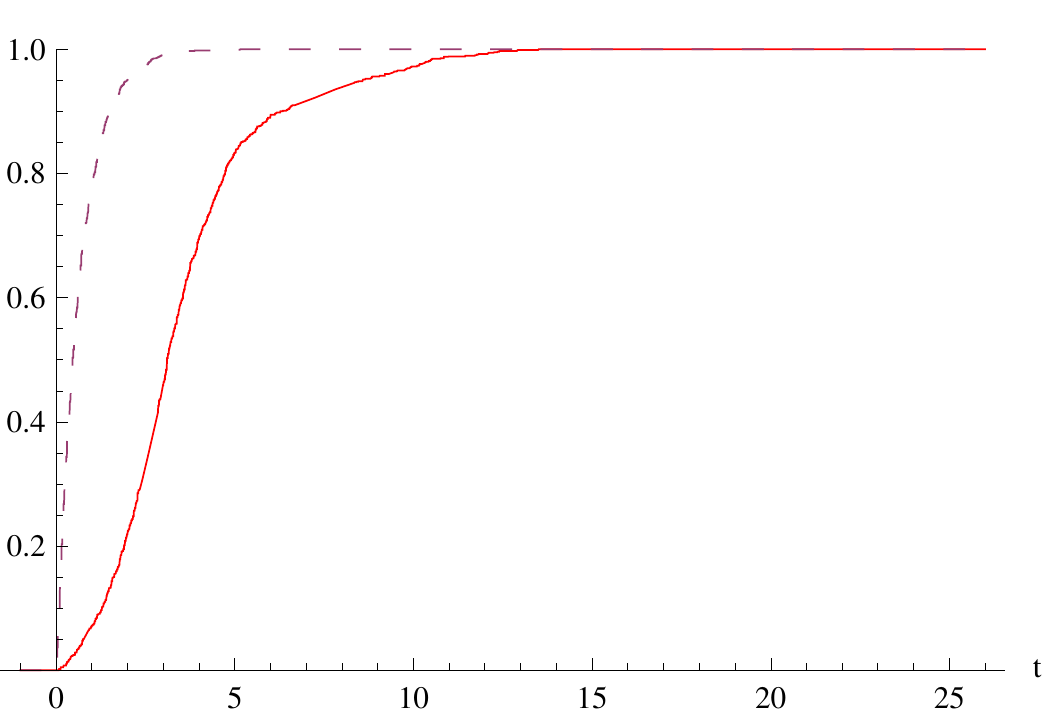}}
\caption{$s=25,\ a=11,\ b=a+2=13$. Conditional cumulative distribution function of the \emph{first passage time} (Frame (a)) and  \emph{sojourn time} (Frame (b)) at site $(a +1,b)$, \emph{given} that in the time interval $(0,s)$ the process visits $(a +1,b)$. Solid line: the interacting case. Dashed line: the free case. Both distribution functions are estimated from the subsample of those trajectories that do hit $(a+1,b)$.}
\label{fig:sogg}
\end{figure}

It is fairly intuitive to attribute the effect shown in \fref{fig:figura1b} to the fact that, in the case $b=a+2$, not only the computational path $\sigma_3 \sigma_1 \sigma_3 \sigma_1=-I$ is available, but also the path $\sigma_3 \sigma_3 \sigma_1 \sigma_1= I$, corresponding to the fact that the rightmost cursor can \emph{wait} in $b$ for the leftmost cursor to jump in $a+1$ ( $\sigma_1 \sigma_1=I$ being thus applied to the register) and \emph{then} both of them can jump to the right of $b$ ($\sigma_3 \sigma_3=I$ being thus applied).\\
The above intuition (involving correlations between positions of the cursors at \emph{different} times) can be made more precise in terms of the stochastic process $(q_1(t), q_2(t))$ associated, according to the prescription of \cite{guerra84} (as specialized to the present context in \cite{defa08}) to the time evolution $\psi_t(x_1,x_2)$, in $H_+=range(P_+)$, of the initial condition \eref{eq:ciperla}. The transition probability per unit time from site $(x_1,x_2)$ to site $(y_1,y_2)$ is given by
\begin{align}
&v_t(y_1,y_2 | x_1,x_2) = \left | h_+((x_1,x_2),(y_1,y_2))\right | \left |\frac{\psi_t(y_1,y_2)}{\psi_t(x_1,x_2)} \right | \cdot \\
&\cdot [ 1+\sin(Arg(\psi_t(x_1,x_2))-Arg(\psi_t(y_1,y_2)) +  Arg(h_+((x_1,x_2),(y_1,y_2))))]. \nonumber
\end{align}
We will indicate by $(q_1^0(t),q_2^0(t)),\ v_t^0(y_1,y_2 | x_1,x_2),\ldots,$ the analogously defined quantities in the absence of interaction. Simulation of processes with the above transition fields are performed, in what follows, according to the first order algorithm outlined in \cite{defa08}.\\
The free process $(q_1^0(t),q_2^0(t))$ (some sample paths of which are shown in the insets of \fref{fig:Xcomp}) can, in the region $\{(x_1,x_2) \in \Lambda_s \times \Lambda_s: 1 \leq x_1 < x_2 \leq s\}$, be described in much the same way as the paradigmatic example of \cite{defa08}: each component starts as a pure birth process (only steps to the right $(x_1 \rightarrow x_1+1)$ or upwards $(x_2 \rightarrow x_2+1)$ are allowed in an initial time interval); at each instant each link (edge between nearest neighbour lattice sites) can be traversed only in one direction; the allowed direction along a link is inverted each time the probability mass at one vertex of the link vanishes.\\ 
Mastering the slalom at $(a +1,b)$ in the interacting case, as shown by \fref{fig:Xcomp}, requires subtle time correlations. There is an instant of time at which most of the trajectories that hit $(a +1,b)$ are there simultaneously. At a later time they radiate from $(a +1,b)$ in many
different directions. As \fref{fig:figura24} shows, given that the process hits $(a +1,b)$, it stays there, on the average, for a longer time in the interacting case than in the free case.
%
%
%
%

%% file: conclusionsandoutlook.tex
\section{Conclusions and outlook}
We have tried, in this note, to contribute to the effort of looking at quantum mechanics as a source of metaphors suggesting Markov processes with \emph{interesting} dynamical behaviour, \emph{interesting} from the point of view of, say, efficiently crossing (in the sense of \cite{farhi97}) a graph or a decision tree, or sampling a function to be minimized \cite{apo89,defa88,santoro06}.\\
In \cite{defa08} we have, in this spirit, shown how to mimic, by a Markov process of the class proposed in \cite{guerra84}, the diffraction effect due to a sharp initial position. Here we have tried to formulate in the same stochastic language the interference effects due to different localizations, as shown in \fref{fig:figura2}, \mbox{of discontinuities of the Hamiltonian}.\\
In the process of doing so we have explored the notions of first passage and sojourn times for a quantum walk, which might prove useful from the point of view of suggesting heuristics
of quantum algorithms, in a context, such as Feynman's \cite{feyn86}, in which timing and synchronization issues play a major role.\\
From the point of view of physics, our analysis raises the question of finding, in the \emph{quantum mechanics} of Anderson localization by non commuting impurities, an analog of the time dependent  phenomenon shown by the \emph{stochastic process} in figures \ref{fig:Xcomp} and \ref{fig:sogg}: the sudden formation (see \fref{fig:figura27}), in the situation of \fref{fig:figura2b}, of a probability bubble at $(a +1,b)$ and its delayed (see \fref{fig:figura24}) bursting in random directions.